\documentclass[conference,romanappendices]{IEEEtran}
\IEEEoverridecommandlockouts

\usepackage{algorithm,algorithmic,amsmath,amssymb,amsthm,bbm,cite,color,graphicx,microtype,setspace,tikz,url}
\usepackage[USenglish]{babel}

\usepackage[utf8]{inputenc}
\usepackage[T1]{fontenc}
\usepackage{hyperref}

\renewcommand{\b}{\mathbf{b}}

\renewcommand{\d}{\mathbf{d}}

\renewcommand{\r}{\mathbf{r}}
\newcommand{\s}{\mathbf{s}}
\renewcommand{\t}{\mathbf{t}}
\renewcommand{\u}{\mathbf{u}}

\newcommand{\x}{\mathbf{x}}
\newcommand{\y}{\mathbf{y}}
\newcommand{\z}{\mathbf{z}}

\newcommand{\0}{\mathbf{0}}

\newcommand{\C}{\mathbf{C}}

\newcommand{\G}{\mathbf{G}}
\renewcommand{\H}{\mathbf{H}}
\newcommand{\I}{\mathbf{I}}

\newcommand{\V}{\mathbf{V}}
\newcommand{\W}{\mathbf{W}}

\newcommand{\setC}{\mathcal{C}}

\newcommand{\setN}{\mathcal{N}}

\newcommand{\setQ}{\mathcal{Q}}

\newcommand{\Compl}{\mbox{$\mathbb{C}$}}

\newcommand{\argmin}{\operatornamewithlimits{argmin}}

\newcommand{\Diag}{\mathrm{Diag}}

\newcommand{\Exp}{\mathbb{E}}
\newcommand{\herm}{\mathrm{H}}
\renewcommand{\Im}{\mathrm{Im}}

\renewcommand{\Re}{\mathrm{Re}}
\newcommand{\sgn}{\mathrm{sgn}}
\newcommand{\tr}{\mathrm{tr}}

\newtheorem{proposition}{Proposition}

\newcommand{\red}[1]{\textcolor{red}{#1}}

\newcommand{\rx}{\textnormal{\tiny{RX}}}
\newcommand{\tx}{\textnormal{\tiny{TX}}}

\usetikzlibrary{arrows, arrows.meta, backgrounds, calc, intersections, decorations.markings, decorations.pathreplacing, positioning, shapes}
\usepackage{pgfplots}
\pgfplotsset{compat=1.17}

\usepackage{titlesec}
\let\subparagraph\relax
\titlespacing{\section}{0pt}{6pt plus 2pt minus 1pt}{4pt plus 1pt minus 1pt} 
\titlespacing{\subsection}{0pt}{4pt plus 2pt minus 1pt}{2pt plus 1pt minus 1pt} 
\usepackage{fancyhdr}
\fancypagestyle{firstpage}{
  
  \fancyhead[R]{{\scriptsize 1}}
  \fancyfoot[C]{{\footnotesize \begin{singlespace} \textcopyright 2023 IEEE. Personal use of this material is permitted. Permission from IEEE must be obtained for all other uses, in any current or future media, including reprinting/republishing this material for advertising or promotional purposes, creating new collective works, for resale or redistribution to servers or lists, or reuse of any copyrighted component of this work in other works. \end{singlespace}}}}

\title{Doubly 1-Bit Quantized Massive MIMO}
\author{
\IEEEauthorblockN{Italo~Atzeni,$^{1}$ Antti~Tölli,$^{1}$ Duy~H.~N.~Nguyen,$^{2}$ and A.~Lee~Swindlehurst$^{3}$} \\ \vspace{-1mm}
\IEEEauthorblockA{$^{1}$ Centre for Wireless Communications, University of Oulu, Finland \\
$^{2}$ Department of Electrical and Computer Engineering, San Diego State University, USA \\
$^{3}$ Department of Electrical Engineering and Computer Science, University of California, Irvine, USA \\
Emails: \{italo.atzeni, antti.tolli\}@oulu.fi, duy.nguyen@sdsu.edu, swindle@uci.edu}
\thanks{This work was supported by the Research Council of Finland (336449 Profi6, 346208 6G~Flagship, 348396 HIGH-6G, and 357504 EETCAMD), by the European Commission (101095759 Hexa-X-II), and by the National Science Foundation (ECCS-2146436, CCF-2225575, and CCF-2225576).} \vspace{-1mm}}

\begin{document}

\maketitle

\thispagestyle{firstpage}

\begin{abstract}
Enabling communications in the (sub-)THz band will call for massive multiple-input multiple-output (MIMO) arrays at either the transmit- or receive-side, or at both. To scale down the complexity and power consumption when operating across massive frequency and antenna dimensions, a sacrifice in the resolution of the digital-to-analog/analog-to-digital converters (DACs/ADCs) will be inevitable. In this paper, we analyze the extreme scenario where both the transmit- and receive-side are equipped with fully digital massive MIMO arrays and 1-bit DACs/ADCs, which leads to a system with minimum radio-frequency complexity, cost, and power consumption. Building upon the Bussgang decomposition, we derive a tractable approximation of the mean squared error (MSE) between the transmitted data symbols and their soft estimates. Numerical results show that, despite its simplicity, a doubly 1-bit quantized massive MIMO system with very large antenna arrays can deliver an impressive performance in terms of MSE and symbol error rate.
\end{abstract}

\begin{IEEEkeywords}
1-bit ADCs, 1-bit DACs, doubly massive MIMO, (sub-)THz communications.
\end{IEEEkeywords}

\section{Introduction} \label{sec:Intro}

Beyond-5G wireless systems are envisioned to exploit the large amount of bandwidth in the sub-THz and THz frequency ranges (0.1--0.3~THz and 0.3--3~THz, respectively) \cite{Raj20}. To overcome the severe pathloss and penetration loss, massive multiple-input multiple-output (MIMO) arrays will be needed at either the transmit- or receive-side, or at both \cite{Buz18}. Fully digital massive MIMO architectures provide highly flexible beamforming and large-scale spatial multiplexing while avoiding the complex beam-management schemes of their hybrid analog-digital counterparts. However, the power consumption of the digital-to-analog/analog-to-digital converters (DACs/ADCs), which scales linearly with the bandwidth and exponentially with the number of resolution bits \cite{Atz21b}, poses significant practical challenges. In this context, low-resolution DACs/ADCs have been considered as a promising enabler of truly massive, fully digital antenna arrays \cite{Sax17,Li17,Jac17}. Simple 1-bit DACs/ADCs can also alleviate the overall complexity and power consumption of the radio-frequency (RF) chains. For instance, 1-bit DACs at the transmitter allow the use of low-cost power amplifiers that are not constrained to operate with backoff \cite{Li21}, whereas 1-bit ADCs at the receiver relax the requirements on the automatic gain control. Hence, the performance loss arising from the 1-bit quantization can be compensated for by adding very low-cost antennas and RF chains.

Massive MIMO systems with low-resolution data converters have been generally studied assuming coarse quantization at the base station (either in the DACs \cite{Sax17,Jac17a} or in the ADCs \cite{Li17,Atz22}) and full-resolution user equipment. In this paper, we analyze the extreme scenario where both the transmit- and receive-side are equipped with fully digital massive MIMO arrays \cite{Buz18} and 1-bit DACs/ADCs, which is referred to in the following as \textit{doubly 1-bit quantized massive MIMO}. Indeed, combining 1-bit DACs and ADCs leads to a fully digital system with minimum RF complexity, cost, and power consumption. Furthermore, the short wavelengths characterizing the (sub-)THz band and the use of 1-bit DACs/ADCs can pave the way for compact and energy-efficient massive MIMO arrays that can also be implemented at the user equipment (e.g., on cars and unmanned aerial vehicles). Only a handful of works in the literature have considered MIMO systems with both 1-bit DACs and 1-bit ADCs. For instance, \cite{Nam19} studied the 1-bit multiple-input single-output capacity with perfect channel state information (CSI) at both the transmitter and receiver. Moreover, \cite{Baz20} described the 1-bit MIMO capacity with imperfect CSI assuming that the number of transmit or receive antennas tends to infinity. Lastly, \cite{Loz21} showed that a proper combination of transmit beamforming and equiprobable signaling allows the system to operate close to the 1-bit MIMO capacity.

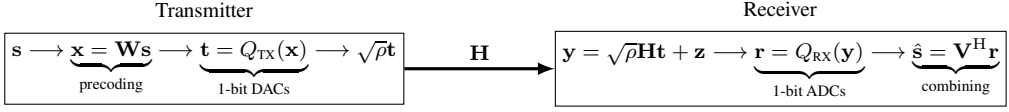
\begin{figure*}
\centering
\begin{tikzpicture}

\footnotesize

\draw node[draw, rectangle, align=center] (tx) {$\s \longrightarrow \underbrace{\x = \W \s}_{\textnormal{precoding}} \longrightarrow \underbrace{\t = Q_{\tx} (\x)}_{\textnormal{1-bit DACs}} \longrightarrow \sqrt{\rho} \t$};

\draw node[draw, rectangle, right=2cm of tx, align=center] (rx) {$\y = \sqrt{\rho} \H \t + \z \longrightarrow \underbrace{\r = Q_{\rx} (\y)}_{\textnormal{1-bit ADCs}} \longrightarrow \underbrace{\hat{\s} = \V^{\herm} \r}_{\textnormal{combining}}$};

\draw node[above=1mm of tx, align=center] () {Transmitter};

\draw node[above=1mm of rx, align=center] () {Receiver};

\begin{scope}[>=latex]
\draw[->, very thick] ($(tx.east)+(0mm,0mm)$) -- ($(rx.west)+(0mm,0mm)$) node[midway, above] {$\H$};
\end{scope}

\end{tikzpicture}
\caption{Diagram of the considered doubly 1-bit quantized massive MIMO system.}
\end{figure*}

The goal of this paper is to provide a tractable analytical framework that lends itself to the performance analysis and optimization of doubly 1-bit quantized massive MIMO systems. Considering a point-to-point system, we build upon the Bussgang decomposition \cite{Dem20} to unfold the relation between the transmitted data symbols (at the input of the transmitter's 1-bit DACs) and the soft-estimated symbols acquired via linear combining of the quantized received signal (at the output of the receiver's 1-bit ADCs). Assuming perfect channel estimation, we derive a tractable approximation of the mean squared error (MSE) between the transmitted data symbols and their soft estimates as well as the combining strategy that minimizes it. This approximation is accurate in the regime of a large number of transmit antennas, which makes the signal at the input of the receiver's 1-bit ADCs approximately Gaussian. Numerical results show that, despite its simplicity, a doubly 1-bit quantized massive MIMO system with very large antenna arrays can deliver an impressive performance in terms of MSE and symbol error rate (SER), which is not far from that of a massive MIMO system with full-resolution DACs and 1-bit ADCs.

\section{System Model} \label{sec:System} \label{sec:2}

We consider a point-to-point doubly 1-bit quantized massive MIMO system, where a transmitter equipped with $N$ antennas and 1-bit DACs transmits $K$ data streams to a receiver with $M$ antennas and 1-bit ADCs, with $K \leq \min (N,M)$. Such a point-to-point system may represent, e.g., a wireless backhaul scenario; however, the following discussion can be easily extended to a multi-user uplink or downlink scenario. To model the 1-bit DACs and ADCs, we introduce the 1-bit quantization function $Q_{\textnormal{\tiny{A}}} (\cdot) : \Compl^{B \times 1} \to \setQ_{\textnormal{\tiny{A}}}$, with $\setQ_{\textnormal{\tiny{A}}} \triangleq \sqrt{\frac{\eta_{\textnormal{\tiny{A}}}}{2}} \{ \pm 1 \pm j \}^{B \times 1}$ and
\begin{align}
Q_{\textnormal{\tiny{A}}} (\b) \triangleq \sqrt{\frac{\eta_{\textnormal{\tiny{A}}}}{2}} \Big( \sgn \big( \Re [\b] \big) + j \, \sgn \big( \Im [\b] \big) \Big) \label{eq:Q}.
\end{align}
Note that the output of the 1-bit quantization function is a vector of scaled quadrature phase-shift keying (QPSK) symbols. In the following, we use the subscripts $\textnormal{A} = \textnormal{TX}$ and $\textnormal{A} = \textnormal{RX}$ to indicate ``transmitter'' and ``receiver'', respectively.

Let $\H \in \Compl^{M \times N}$ denote the channel matrix between the transmitter and the receiver. In this paper, we assume that $\H$ is perfectly known at both the transmitter and receiver, and we leave the analysis with imperfect CSI for future work. The transmitter aims at conveying the data symbol vector $\s \in \Compl^{K \times 1}$ to the receiver. As in \cite{Sax17,Jac17a}, we consider a quantized linear precoding strategy, whereby the precoding matrix is designed based on $\H$ and independently of $\s$ and the subsequent quantization step.\footnote{An alternative approach, which goes beyond the scope of this paper, is symbol-level precoding, whereby the analog signal at the output of the 1-bit DACs is designed based on $\H$ and $\s$ \cite{Li21}. Symbol-level precoding outperforms its quantized linear counterpart at the cost of higher complexity.} In this setting, $\s$ is first precoded as
\begin{align}
\x \triangleq \W \s \in \Compl^{N \times 1} \label{eq:x}
\end{align}
where $\W \in \Compl^{N \times K}$ is the precoding matrix, and then quantized at the 1-bit DACs as
\begin{align}
\t \triangleq Q_{\tx} (\x). \label{eq:t}
\end{align}
Here, the scaling factor $\eta_{\tx}$ of the 1-bit quantization function in \eqref{eq:Q} is fixed as $\eta_{\tx} = \frac{1}{N}$ to satisfy the power constraint $\| \t \|^{2} = 1$.

Subsequently, the analog signal $\t$ is transmitted over the channel with transmit power $\rho$ and the signal arriving at the receiver is given by
\begin{align}
\y \triangleq \sqrt{\rho} \H \t + \z \in \Compl^{M \times 1} \label{eq:y}
\end{align}
where $\z \sim \setC \setN (\0, \I_{M})$ is a vector of additive white Gaussian noise (AWGN). Since the AWGN has unit variance, $\rho$ can be interpreted as the transmit signal-to-noise ratio (SNR). Then, $\y$ is quantized at the 1-bit ADCs as
\begin{align}
\r \triangleq Q_{\rx} (\y) = Q_{\rx} \big( \sqrt{\rho} \H Q_{\tx} (\W \s) + \z \big). \label{eq:r}
\end{align}
Here, the scaling factor $\eta_{\rx}$ of the 1-bit quantization function in \eqref{eq:Q} is fixed as $\eta_{\rx} = \rho + 1$ so that the element-wise variance of the output coincides with that of the input when the channel elements have unit variance and $N \to \infty$ (see \eqref{eq:C_y} in the following). Note that the doubly 1-bit quantized signal in \eqref{eq:r} is what is observed at the receiver. Finally, the receiver acquires a soft estimate of $\s$ via linear combining of the digital signal $\r$ as
\begin{align}
\hat{\s} \triangleq \V^{\herm} \r \in \Compl^{K \times 1} \label{eq:s_hat}
\end{align}
where $\V \in \Compl^{M \times K}$ is the combining matrix.

In this paper, we consider the MSE between $\s$ and its soft estimate $\hat{\s}$ in \eqref{eq:s_hat} as a performance metric, which is given by
\begin{align}
\varepsilon \triangleq \frac{1}{K} \Exp_{\s, \z} [ \| \s - \hat{\s} \|^{2}]. \label{eq:MSE}
\end{align}
We observe that deriving the above MSE requires obtaining a tractable expression for $\r$ in \eqref{eq:r} as a function of $\s$.

\section{Linearization Via the Bussgang Decomposition}

In this section, we consider Gaussian data symbols, i.e., $\s \sim \setC \setN (\0, \I_{K})$. Then, we express the doubly 1-bit quantized signal observed at the receiver as a linear function using the Bussgang decomposition \cite{Dem20}, which allows one to write the output of a nonlinear system as a scaled version of the input plus an uncorrelated distortion. Finally, we derive a tractable approximation of the MSE in \eqref{eq:MSE}.

\subsection{Linearization at the Transmitter} \label{sec:3_1}

\begin{figure*}
\begin{align}
\addtocounter{equation}{+6}
\C_{\t} & = \frac{2}{\pi} \eta_{\tx} \Big( \arcsin \big( \Diag (\C_{\x})^{- \frac{1}{2}} \Re [\C_{\x}] \Diag (\C_{\x})^{- \frac{1}{2}} \big) + j \, \arcsin \big( \Diag (\C_{\x})^{- \frac{1}{2}} \Im [\C_{\x}] \Diag (\C_{\x})^{- \frac{1}{2}} \big) \Big), \label{eq:C_t_2} \\
\addtocounter{equation}{+6}
\tilde{\C}_{\r} & \triangleq \frac{2}{\pi} \eta_{\rx} \Big( \arcsin \big( \Diag (\C_{\y})^{- \frac{1}{2}} \Re [\C_{\y}] \Diag (\C_{\y})^{- \frac{1}{2}} \big) + j \, \arcsin \big( \Diag (\C_{\y})^{- \frac{1}{2}} \Im [\C_{\y}] \Diag (\C_{\y})^{- \frac{1}{2}} \big) \Big) \label{eq:C_r_tilde}
\end{align}
\addtocounter{equation}{-9}
\vspace{-2mm}
\hrulefill
\vspace{-2mm}
\end{figure*}

Let us define
\addtocounter{equation}{-5}
\begin{align}
\C_{\x} & \triangleq \Exp_{\s} [\x \x^{\herm}] = \W \W^{\herm} \in \Compl^{N \times N}, \label{eq:C_x} \\
\C_{\t} & \triangleq \Exp_{\s} [\t \t^{\herm}] \in \Compl^{N \times N}. \label{eq:C_t}
\end{align}
Then, we use the Bussgang decomposition to linearize $\t$ in \eqref{eq:t} with respect to $\x$ in \eqref{eq:x} (and, thus, with respect to $\s$) as
\begin{align}
\t = \G_{\tx} \x + \d_{\tx} \label{eq:t_Buss}
\end{align}
where $\d_{\tx} \in \Compl^{N \times 1}$ is a zero-mean, non-Gaussian distortion vector that is uncorrelated with $\x$ (and, obviously, with $\s$) and
\begin{align}
\G_{\tx} \triangleq \Exp_{\s} [\t \x^{\herm}] \C_{\x}^{-1} \in \Compl^{N \times N} \label{eq:G_tx}
\end{align}
is the Bussgang gain matrix. Indeed, $\G_{\tx} \x$ and $\d_{\tx}$ are the minimum MSE estimate of $\t$ given $\x$ and the corresponding estimation error, respectively. Moreover, according to \eqref{eq:t_Buss}, $\C_{\t}$ in \eqref{eq:C_t} can be obtained as
\begin{align}
\C_{\t} = \G_{\tx} \C_{\x} \G_{\tx} + \C_{\d_{\tx}} \label{eq:C_t_1}
\end{align}
with $\C_{\d_{\tx}} \triangleq \Exp [\d_{\tx} \d_{\tx}^{\herm}] \in \Compl^{N \times N}$.

Since $\s$ is Gaussian, we have that $\G_{\tx}$ is a diagonal matrix that can be computed in closed form following well-known steps (see, e.g., \cite{Li17}) as
\begin{align}
\G_{\tx} = \sqrt{\frac{2}{\pi} \eta_{\tx}} \Diag (\C_{\x})^{- \frac{1}{2}}. \label{eq:G_tx_1}
\end{align}
Note that, when $K < N$ (i.e., when the number of data streams is strictly smaller than the number of transmit antennas, which is generally the case in downlink massive MIMO systems), $\C_{\x}$ in \eqref{eq:C_x} is rank-deficient and $\G_{\tx}$ cannot be computed as in \eqref{eq:G_tx};\footnote{When $\C_{\x}$ is rank-deficient, its inverse in \eqref{eq:G_tx} can be replaced by its pseudoinverse, as described in \cite{Bjo19}. In this case, the resulting $\G_{\tx}$ may not be diagonal, although $\G_{\tx} \x$ will be the same whether $\G_{\tx}$ is obtained via \eqref{eq:G_tx} or \eqref{eq:G_tx_1}.} nonetheless, it can still be obtained via \eqref{eq:G_tx_1}, which does not involve the inversion of $\C_{\x}$. Furthermore, $\C_{\t}$ in \eqref{eq:C_t} can be computed in closed form as in \eqref{eq:C_t_2} at the top of the page \cite{Li17}.

\subsection{Linearization at the Receiver} \label{sec:3_2}

Let us define
\addtocounter{equation}{+1}
\begin{align}
\C_{\y} & \triangleq \Exp_{\s,\z} [\y \y^{\herm}] = \rho \H \C_{\t} \H^{\herm} + \I_{M} \in \Compl^{M \times M}, \label{eq:C_y} \\
\C_{\r} & \triangleq \Exp_{\s,\z} [\r \r^{\herm}] \in \Compl^{M \times M} \label{eq:C_r}
\end{align}
where we point out that the above expectations are taken also over the AWGN. Then, we use the Bussgang decomposition to linearize $\r$ in \eqref{eq:r} with respect to $\y$ in \eqref{eq:y} as
\begin{align}
\r = \G_{\rx} \y + \d_{\rx} = \G_{\rx} \big( \sqrt{\rho} \H (\G_{\tx} \W \s + \d_{\tx}) + \z \big) + \d_{\rx} \label{eq:r_Buss}
\end{align}
where $\d_{\rx} \in \Compl^{M \times 1}$ is a zero-mean, non-Gaussian distortion vector that is uncorrelated with $\y$ and
\begin{align}
\G_{\rx} \triangleq \Exp_{\y} [\r \y^{\herm}] \C_{\y}^{-1} \in \Compl^{M \times M} \label{eq:G_rx}
\end{align}
is the Bussgang gain matrix. Moreover, according to \eqref{eq:r_Buss}, $\C_{\r}$ in \eqref{eq:C_r} can be obtained as
\begin{align}
\C_{\r} \triangleq \G_{\rx} \C_{\y} \G_{\rx} + \C_{\d_{\rx}} \label{eq:C_r_1}
\end{align}
with $\C_{\d_{\rx}} \triangleq \Exp [\d_{\rx} \d_{\rx}^{\herm}] \in \Compl^{M \times M}$.

At this stage, we have achieved our goal to express the doubly 1-bit quantized signal observed at the receiver as a linear function of $\s$. However, since $\y$ is \textit{not} Gaussian due to the 1-bit DACs at the transmitter,\footnote{This can be also observed from \eqref{eq:r_Buss}, which includes the non-Gaussian distortion vector $\d_{\tx}$.} we have that $\G_{\rx}$ is generally not diagonal. Consequently, $\C_{\r}$ and $\G_{\rx}$ are not available in closed form, which makes the expression in \eqref{eq:r_Buss} not tractable. Despite that, we make the following observation: each element of $\u \triangleq \H \t \in \Compl^{M \times 1}$ is a weighted sum of $N$ scaled QPSK symbols. Therefore, when the number of transmit antennas $N$ is large and the channel elements are weakly correlated, we have that $\u$ is approximately Gaussian, i.e., $\u \mathrel{\dot{\sim}} \setC \setN (\0, \H \C_{\t} \H^{\herm})$. This approximation becomes asymptotically exact as $N \to \infty$; nonetheless, with i.i.d. channel elements, we observe through numerical simulations that it is already quite accurate for $N \geq 8$. In this context, we have that $\y$ is approximately Gaussian, i.e., $\y \mathrel{\dot{\sim}} \setC \setN (\0, \C_{\y})$. Exploiting this argument, $\G_{\rx}$ in \eqref{eq:G_rx} can be approximated as $\G_{\rx} \approx \tilde{\G}_{\rx}$, where
\begin{align}
\tilde{\G}_{\rx} \triangleq \sqrt{\frac{2}{\pi} \eta_{\rx}} \Diag (\C_{\y})^{- \frac{1}{2}} \label{eq:G_rx_tilde}
\end{align}
is obtained following similar steps as for \eqref{eq:G_tx_1}. Furthermore, $\C_{\r}$ in \eqref{eq:C_r} can be approximated as $\C_{\r} \approx \tilde{\C}_{\r}$, where $\tilde{\C}_{\r}$ is given in \eqref{eq:C_r_tilde} at the top of the page and results from following similar steps as for \eqref{eq:C_t_2}.

\subsection{Tractable Approximation of the MSE} \label{sec:3_3}

Building upon the linearizations at the transmitter and at the receiver described in Sections~\ref{sec:3_1} and~\ref{sec:3_2}, respectively, we approximate the MSE in \eqref{eq:MSE} as follows.
\begin{proposition} \label{pro:MSE}
The MSE in \eqref{eq:MSE} can be approximated as $\varepsilon \approx \tilde{\varepsilon}$, with
\addtocounter{equation}{+1}
\begin{align}
\tilde{\varepsilon} \triangleq 1 + \frac{1}{K} \tr (\V^{\herm} \tilde{\C}_{\r} \V) - \frac{2}{K} \sqrt{\rho} \tr \big( \Re [\V^{\herm} \tilde{\G}_{\rx} \H \G_{\tx} \W] \big). \label{eq:MSE_tilde}
\end{align}
\end{proposition}

\begin{IEEEproof}
See Appendix~\ref{app:1}
\end{IEEEproof}

\medskip

\noindent Our numerical simulations in Section~\ref{sec:4} show that the approximate MSE $\tilde{\varepsilon}$ in \eqref{eq:MSE_tilde} behaves as an upper bound on the true MSE $\varepsilon$. Moreover, for a fixed precoding matrix $\W$,\footnote{Note that \eqref{eq:MSE_tilde} depends on the specific choice of the precoding matrix $\W$ and also on $\C_{\x}$ in \eqref{eq:C_x} through $\G_{\tx}$ in \eqref{eq:G_tx_1}, $\tilde{\G}_{\rx}$ in \eqref{eq:G_rx_tilde}, and $\tilde{\C}_{\r}$ in \eqref{eq:C_r_tilde}.} we have that $\tilde{\varepsilon}$ is a convex quadratic function of the combining matrix $\V$. Hence, we can minimize the approximate MSE with respect to $\V$, which yields
\begin{align}
\V^{\star} \triangleq \underset{\V}{\argmin} \, \tilde{\varepsilon} = \sqrt{\rho} \tilde{\C}_{\r}^{-1} \tilde{\G}_{\rx} \H \G_{\tx} \W. \label{eq:V_star}
\end{align}

\section{Numerical Results} \label{sec:4}

In this section, we evaluate the performance of the considered doubly 1-bit quantized massive MIMO system in terms of MSE between the transmitted data symbols and their soft estimates. We consider far-field propagation and generate the channel matrix $\H$ using the discrete physical channel model \cite{Say02}. The transmitter and receiver are equipped with square uniform planar arrays of $\sqrt{N} \times \sqrt{N}$ (resp. $\sqrt{M} \times \sqrt{M}$) half-wavelength spaced antennas and are placed with their broadsides facing each other. Between them lies a cluster of $10^{2}$ scatterers, which give rise to as many independent propagation paths. For both the transmitter and receiver, the scatterers are confined within an angle spread of $\frac{\pi}{6}$ around the broadside direction in both the azimuth and elevation. The channels are normalized such that their elements have unit variance and the pathloss is incorporated into the transmit SNR $\rho$. The following numerical results are obtained by averaging over $10^{3}$ independent channel realizations. For each realization of $\H$, the precoding matrix $\W$ comprises the $K$ principal right eigenvectors of $\H$, whereas the combining matrix $\V$ is computed as in \eqref{eq:V_star}, which minimizes the approximate MSE in \eqref{eq:MSE_tilde}. The values of the simulation parameters $N$, $M$, $K$, and $\rho$ are reported above each figure.

\begin{figure}
\centering
\begin{tikzpicture}

\begin{axis}[
	width=8cm,
	height=6.5cm,
	xmin=400, xmax=1600,
	ymin=0, ymax=0.2,
	xlabel={$N = M$},
	ylabel={MSE},
	xtick={400,576,784,1024,1296,1600},
	ytick={0,0.05,0.1,0.15,0.2},
	xticklabels={400,576,784,1024,1296,1600},
	yticklabels={0,0.05,0.1,0.15,0.2},
	xlabel near ticks,
	ylabel near ticks,
	x label style={font=\footnotesize},
	y label style={font=\footnotesize},
	ticklabel style={font=\footnotesize},
	legend style={at={(0.98,0.98)}, anchor=north east},
	legend style={font=\scriptsize, inner sep=1pt, fill opacity=0.75, draw opacity=1, text opacity=1},
	legend cell align=left,
	grid=both,
	title={$K = 16$, $\rho = 10$~dB},
	title style={font=\scriptsize},
]

\addplot[thick, black]
table [x=N, y=MSE_tilde_2, col sep=comma] {Figures/files_txt/MSE_vs_N=M_K=16_rho=10dB.txt};
\addlegendentry{$\tilde{\varepsilon}$ (1-bit DACs/ADCs)};

\addplot[thick, black, only marks, mark=asterisk]
table [x=N, y=MSE_mc_2, col sep=comma] {Figures/files_txt/MSE_vs_N=M_K=16_rho=10dB.txt};
\addlegendentry{$\varepsilon$ (1-bit DACs/ADCs)};

\addplot[thick, cyan, dashed]
table [x=N, y=MSE_1, col sep=comma] {Figures/files_txt/MSE_vs_N=M_K=16_rho=10dB.txt};
\addlegendentry{$\varepsilon$ (1-bit ADCs)};

\end{axis}

\end{tikzpicture}

\caption{MSE for 1-bit DACs/ADCs and full-resolution DACs/1-bit ADCs versus number of transmit/receive antennas. The true MSE $\varepsilon$ is obtained via Monte Carlo simulations.}
\label{fig:MSE_vs_NM}
\end{figure}
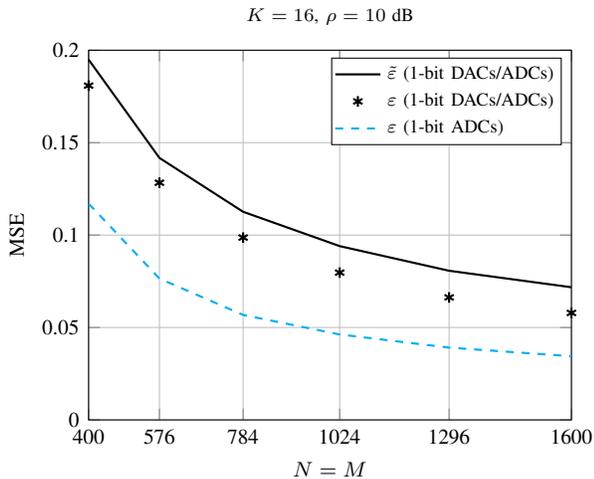

\begin{figure}
\centering
\begin{tikzpicture}

\begin{axis}[
	width=8cm,
	height=6.5cm,
	xmin=400, xmax=1600,
	ymin=0.05, ymax=0.2,
	xlabel={$N$},
	ylabel={$\tilde{\varepsilon}$},
	xtick={400,576,784,1024,1296,1600},
	ytick={0.05,0.1,0.15,0.2},
	xticklabels={400,576,784,1024,1296,1600},
	yticklabels={0.05,0.1,0.15,0.2},
	xlabel near ticks,
	ylabel near ticks,
	x label style={font=\footnotesize},
	y label style={font=\footnotesize},
	ticklabel style={font=\footnotesize},
	legend style={at={(0.98,0.98)}, anchor=north east},
	legend style={font=\scriptsize, inner sep=1pt, fill opacity=0.75, draw opacity=1, text opacity=1},
	legend cell align=left,
	grid=both,
	xmode=log,
	log basis x={2},
	title={$K = 16$, $\rho = 10$~dB},
	title style={font=\scriptsize},
]

\addplot[thick, red]
table [x=N, y=MSE_tilde_2, col sep=comma] {Figures/files_txt/MSE_vs_N_M=400_K=16_rho=10dB.txt};
\addlegendentry{$M = 400$};

\addplot[thick, blue]
table [x=N, y=MSE_tilde_2, col sep=comma] {Figures/files_txt/MSE_vs_N_M=1024_K=16_rho=10dB.txt};
\addlegendentry{$M = 1024$};

\addplot[thick, green]
table [x=N, y=MSE_tilde_2, col sep=comma] {Figures/files_txt/MSE_vs_N_M=1600_K=16_rho=10dB.txt};
\addlegendentry{$M = 1600$};

\addplot[thick, red, densely dotted]
table [x=M, y=MSE_tilde_2, col sep=comma] {Figures/files_txt/MSE_vs_M_N=400_K=16_rho=10dB.txt};

\addplot[thick, blue, densely dotted]
table [x=M, y=MSE_tilde_2, col sep=comma] {Figures/files_txt/MSE_vs_M_N=1024_K=16_rho=10dB.txt};

\addplot[thick, green, densely dotted]
table [x=M, y=MSE_tilde_2, col sep=comma] {Figures/files_txt/MSE_vs_M_N=1600_K=16_rho=10dB.txt};
\end{axis}

\end{tikzpicture}
\caption{Approximate MSE for 1-bit DACs/ADCs versus number of transmit antennas for different numbers of receive antennas. The dotted curves are obtained by switching $N$ and $M$.}
\label{fig:MSE_vs_N}
\end{figure}
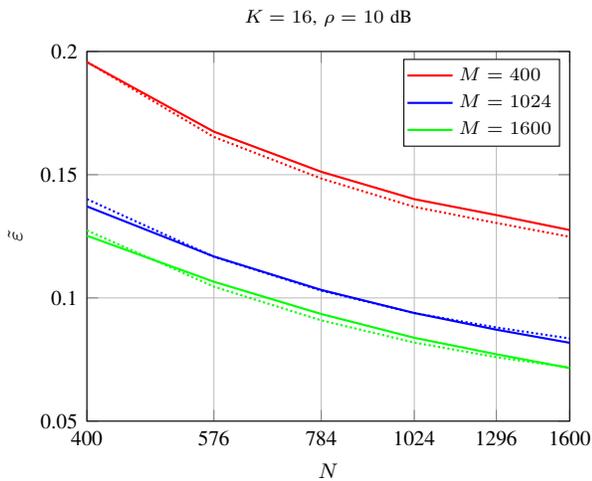

Figure~\ref{fig:MSE_vs_NM} plots the MSE versus the number of transmit/receive antennas, with $N = M$. The approximate MSE $\tilde{\varepsilon}$ in \eqref{eq:MSE_tilde} behaves as a tight upper bound on the true MSE $\varepsilon$ in \eqref{eq:MSE}, where the latter is obtained via Monte Carlo simulations with $10^{3}$ independent realizations of the data symbol vector $\s$ for each realization of $\H$. Furthermore, we observe that truly massive antenna arrays at both the transmitter and receiver are necessary to achieve impressive values of the MSE, e.g., about $5 \times 10^{-2}$ for $N = M = 1600$. Nonetheless, the performance of the considered doubly 1-bit quantized massive MIMO system is not far from that of a massive MIMO system with full-resolution DACs and 1-bit ADCs. Specifically, replacing the 1-bit DACs with full-resolution ones reduces the MSE not even by a factor of two at the cost of much higher RF complexity and power consumption at the transmitter.

\begin{figure}
\centering
\begin{tikzpicture}

\begin{axis}[
	width=8cm,
	height=6.5cm,
	xmin=2, xmax=64,
	ymin=0, ymax=0.5,
	xlabel={$K$},
	ylabel={$\tilde{\varepsilon}$},
	ytick={0,0.1,0.2,0.3,0.4,0.5},
	xticklabels={2,4,8,16,32,64},
	xlabel near ticks,
	ylabel near ticks,
	x label style={font=\footnotesize},
	y label style={font=\footnotesize},
	ticklabel style={font=\footnotesize},
	legend style={at={(0.02,0.98)}, anchor=north west},
	legend style={font=\scriptsize, inner sep=1pt, fill opacity=0.75, draw opacity=1, text opacity=1},
	legend cell align=left,
	grid=both,
	xmode=log,
	log basis x={2},
	title={$\rho = 10$~dB},
	title style={font=\scriptsize},
]

\addplot[thick, red]
table [x=K, y=MSE_tilde_2, col sep=comma] {Figures/files_txt/MSE_vs_K_N=M=400_rho=10dB.txt};
\addlegendentry{$N = M = 400$};

\addplot[thick, blue]
table [x=K, y=MSE_tilde_2, col sep=comma] {Figures/files_txt/MSE_vs_K_N=M=1024_rho=10dB.txt};
\addlegendentry{$N = M = 1024$};

\addplot[thick, green]
table [x=K, y=MSE_tilde_2, col sep=comma] {Figures/files_txt/MSE_vs_K_N=M=1600_rho=10dB.txt};
\addlegendentry{$N = M = 1600$};

\end{axis}

\end{tikzpicture}
\caption{Approximate MSE for 1-bit DACs/ADCs versus number of data streams for different numbers of transmit/receive antennas.}
\label{fig:MSE_vs_K}
\end{figure}
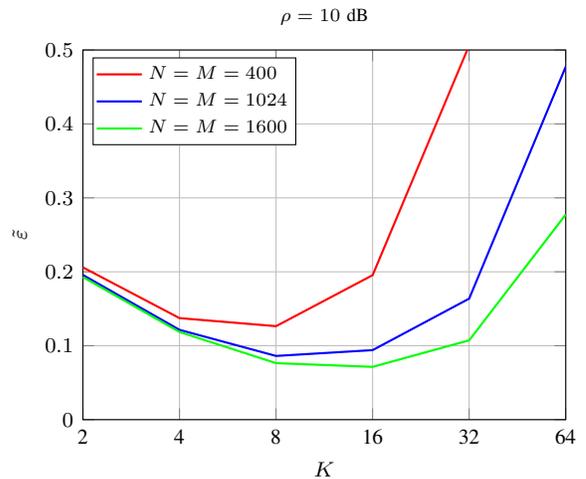

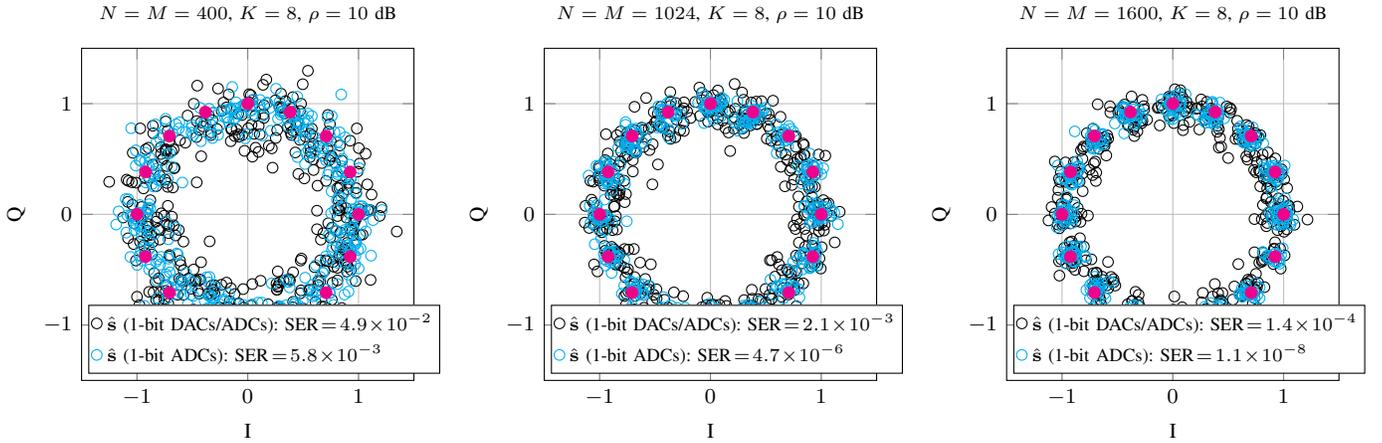
\begin{figure*}
\centering
\hspace{-3mm}
\begin{minipage}{0.32\textwidth}
\begin{tikzpicture}

\begin{axis}[
	width=6cm,
	height=6cm,
	xmin=-1.5, xmax=1.5,
	ymin=-1.5, ymax=1.5,
	xlabel={I},
	ylabel={Q},
	xtick={-1,0,1},
	ytick={-1,0,1},
	xlabel near ticks,
	ylabel near ticks,
	x label style={font=\footnotesize},
	y label style={font=\footnotesize},
	ticklabel style={font=\footnotesize},
	legend style={at={(0.02,0.02)}, anchor=south west},
	legend style={font=\scriptsize, inner sep=1pt, fill opacity=0.75, draw opacity=1, text opacity=1},
	legend cell align=left,
	grid=both,
	title={$N = M = 400$, $K = 8$, $\rho = 10$~dB},
	title style={font=\scriptsize},
]

\addplot[thin, black, only marks, mark=o, mark options={scale=1}]
table[x=s_hat_2_real, y=s_hat_2_imag, col sep=comma] {Figures/files_txt/16PSK_N=M=400_K=8_rho=10dB_2.txt};
\addlegendentry{$\hat{\s}$ (1-bit DACs/ADCs): $\textrm{SER} \! = \! 4.9 \! \times \! 10^{-2}$};

\addplot[thin, cyan, only marks, mark=o, mark options={scale=1}]
table[x=s_hat_1_real, y=s_hat_1_imag, col sep=comma] {Figures/files_txt/16PSK_N=M=400_K=8_rho=10dB_2.txt};
\addlegendentry{$\hat{\s}$ (1-bit ADCs): $\textrm{SER} \! = \! 5.8 \! \times \! 10^{-3}$};

\addplot[thick, magenta, only marks, mark=*, mark options={scale=1}]
table[x=s_psk_real, y=s_psk_imag, col sep=comma] {Figures/files_txt/16PSK_N=M=400_K=8_rho=10dB_1.txt};

\end{axis}

\end{tikzpicture}
\end{minipage}
\hspace{1mm}
\begin{minipage}{0.32\textwidth}
\begin{tikzpicture}

\begin{axis}[
	width=6cm,
	height=6cm,
	xmin=-1.5, xmax=1.5,
	ymin=-1.5, ymax=1.5,
	xlabel={I},
	ylabel={Q},
	xtick={-1,0,1},
	ytick={-1,0,1},
	xlabel near ticks,
	ylabel near ticks,
	x label style={font=\footnotesize},
	y label style={font=\footnotesize},
	ticklabel style={font=\footnotesize},
	legend style={at={(0.02,0.02)}, anchor=south west},
	legend style={font=\scriptsize, inner sep=1pt, fill opacity=0.75, draw opacity=1, text opacity=1},
	legend cell align=left,
	grid=both,
	title={$N = M = 1024$, $K = 8$, $\rho = 10$~dB},
	title style={font=\scriptsize},
]

\addplot[thin, black, only marks, mark=o, mark options={scale=1}]
table[x=s_hat_2_real, y=s_hat_2_imag, col sep=comma] {Figures/files_txt/16PSK_N=M=1024_K=8_rho=10dB_2.txt};
\addlegendentry{$\hat{\s}$ (1-bit DACs/ADCs): $\textrm{SER} \! = \! 2.1 \! \times \! 10^{-3}$};

\addplot[thin, cyan, only marks, mark=o, mark options={scale=1}]
table[x=s_hat_1_real, y=s_hat_1_imag, col sep=comma] {Figures/files_txt/16PSK_N=M=1024_K=8_rho=10dB_2.txt};
\addlegendentry{$\hat{\s}$ (1-bit ADCs): $\textrm{SER} \! = \! 4.7 \! \times \! 10^{-6}$};

\addplot[thick, magenta, only marks, mark=*, mark options={scale=1}]
table[x=s_psk_real, y=s_psk_imag, col sep=comma] {Figures/files_txt/16PSK_N=M=1024_K=8_rho=10dB_1.txt};

\end{axis}

\end{tikzpicture}
\end{minipage}
\hspace{1mm}
\begin{minipage}{0.32\textwidth}
\begin{tikzpicture}

\begin{axis}[
	width=6cm,
	height=6cm,
	xmin=-1.5, xmax=1.5,
	ymin=-1.5, ymax=1.5,
	xlabel={I},
	ylabel={Q},
	xtick={-1,0,1},
	ytick={-1,0,1},
	xlabel near ticks,
	ylabel near ticks,
	x label style={font=\footnotesize},
	y label style={font=\footnotesize},
	ticklabel style={font=\footnotesize},
	legend style={at={(0.02,0.02)}, anchor=south west},
	legend style={font=\scriptsize, inner sep=1pt, fill opacity=0.75, draw opacity=1, text opacity=1},
	legend cell align=left,
	grid=both,
	title={$N = M = 1600$, $K = 8$, $\rho = 10$~dB},
	title style={font=\scriptsize},
]

\addplot[thin, black, only marks, mark=o, mark options={scale=1}]
table[x=s_hat_2_real, y=s_hat_2_imag, col sep=comma] {Figures/files_txt/16PSK_N=M=1600_K=8_rho=10dB_2.txt};
\addlegendentry{$\hat{\s}$ (1-bit DACs/ADCs): $\textrm{SER} \! = \! 1.4 \! \times \! 10^{-4}$};

\addplot[thin, cyan, only marks, mark=o, mark options={scale=1}]
table[x=s_hat_1_real, y=s_hat_1_imag, col sep=comma] {Figures/files_txt/16PSK_N=M=1600_K=8_rho=10dB_2.txt};
\addlegendentry{$\hat{\s}$ (1-bit ADCs): $\textrm{SER} \! = \! 1.1 \! \times \! 10^{-8}$};

\addplot[thick, magenta, only marks, mark=*, mark options={scale=1}]
table[x=s_psk_real, y=s_psk_imag, col sep=comma] {Figures/files_txt/16PSK_N=M=1600_K=8_rho=10dB_1.txt};

\end{axis}

\end{tikzpicture}
\end{minipage}
\caption{Soft-estimated symbols for 1-bit DACs/ADCs and full-resolution DACs/1-bit ADCs with 16-PSK data symbols for different numbers of transmit/receive antennas. The magenta solid marks represent the transmitted data symbols.}
\label{fig:scatter}
\end{figure*}

Figure~\ref{fig:MSE_vs_N} illustrates the approximate MSE versus the number of transmit antennas for different numbers of receive antennas (solid lines) and versus the number of receive antennas for different numbers of transmit antennas (dotted lines). We observe that increasing the number of either transmit or receive antennas produces roughly the same effect. However, since the combining matrix is optimized for a given channel and precoding matrix, the second option provides slightly better results in this case. Figure~\ref{fig:MSE_vs_K} depicts the approximate MSE versus the number of data streams for different numbers of transmit/receive antennas, with $N = M$. For each configuration, there is an optimal number of data streams: on the one hand, judiciously increasing the number of data streams generates a useful scrambling of the 1-bit quantized signals at the $M$ receive antennas \cite{Atz22}; on the other hand, the inter-stream interference becomes dominant for large values of $K$.

\section{Data Detection} \label{sec:5}

In this section, we briefly evaluate the data detection performance with non-Gaussian data symbols in terms of SER. In this respect, we point out that $\x$ in \eqref{eq:x} may be approximately Gaussian even with non-Gaussian data symbols when $K$ is large. Figure~\ref{fig:scatter} plots the soft-estimated symbols with 16-PSK (phase-shift keying) data symbols for different numbers of transmit/receive antennas, with $N = M$. As the number of antennas increases, the dispersion of the soft-estimated symbols around the transmitted data symbols reduces noticeably, which translates into an improved SER performance. For $N = M = 1600$, a remarkable SER in the order of $10^{-4}$ is obtained. Nonetheless, an acceptable SER (considering the absence of coding) in the order of $10^{-2}$ is achieved already for $N = M = 400$.

\section{Conclusions} \label{sec:6}

Among fully digital systems, doubly 1-bit quantized massive MIMO systems are endowed with minimum RF complexity, cost, and power consumption. In this setting, we derived a tractable approximation of the MSE between the transmitted data symbols and their soft estimates as well as the combining strategy that minimizes it. We showed that, despite its simplicity, a doubly 1-bit quantized massive MIMO system with very large antenna arrays can deliver an impressive performance in terms of MSE and SER, which is not far from that of a massive MIMO system with full-resolution DACs and 1-bit ADCs. Future work will analyze the impact of imperfect CSI along with the overall energy efficiency.

\appendices

\section{Proof of Proposition~\ref{pro:MSE}} \label{app:1}

We begin by writing \eqref{eq:MSE} as
\begin{align}
\varepsilon & = \frac{1}{K} \Exp_{\s,\z} \big[ \| \s - \V^{\herm} \r \|^{2} \big] \\
& = \frac{1}{K} \Exp_{\s,\z} \big[ \tr (\s \s^{\herm}) + \tr (\V^{\herm} \r \r^{\herm} \V) - 2 \tr \big( \Re [\V^{\herm} \r \s^{\herm}] \big) \big] \\
& = 1 + \frac{1}{K} \tr (\V^{\herm} \C_{\r} \V) - \frac{2}{K} \tr \big( \Re \big[ \V^{\herm} \Exp [\r \s^{\herm}] \big] \big). \label{eq:der1}
\end{align}
Then, by plugging \eqref{eq:r_Buss} into the remaining expectation term in \eqref{eq:der1}, we obtain
\begin{align}
\Exp [\r \s^{\herm}] & = \Exp_{\s,\z} \Big[ \big( \G_{\rx} \big( \sqrt{\rho} \H (\G_{\tx} \W \s + \d_{\tx}) + \z \big) + \d_{\rx} \big) \s^{\herm} \Big] \\
& = \sqrt{\rho} \G_{\rx} \H \G_{\tx} \W \Exp_{\s} [\s \s^{\herm}] \label{eq:der2}
\end{align}
where \eqref{eq:der2} follows from the fact that $\d_{\tx}$, $\z$, and $\d_{\rx}$ are all uncorrelated with $\s$. Hence, plugging \eqref{eq:der2} into \eqref{eq:der1} yields
\begin{align}
\varepsilon = 1 + \frac{1}{K} \tr (\V^{\herm} \C_{\r} \V) - \frac{2}{K} \sqrt{\rho} \tr \big( \Re [\V^{\herm} \G_{\rx} \H \G_{\tx} \W] \big).
\end{align}
Finally, the approximate MSE in \eqref{eq:MSE_tilde} is obtained by replacing $\G_{\rx}$ with $\tilde{\G}_{\rx}$ in \eqref{eq:G_rx_tilde} and $\C_{\r}$ with $\tilde{\C}_{\r}$ in \eqref{eq:C_r_tilde}. \hfill \IEEEQED

\addcontentsline{toc}{chapter}{References}
\bibliographystyle{IEEEtran}
\bibliography{refs_abbr,refs}

\end{document}